**New Title:** Bridging observational studies and randomized experiments by embedding the former in the latter

**Author names and affiliations:** Marie-Abele C. Bind[1*] and Donald B. Rubin[1]
[1] Department of Statistics, Faculty of Arts and Sciences, Harvard University, Cambridge, MA, USA

**Corresponding author:**
Marie-Abele Bind
Department of Statistics
Science Center, 7th Floor
One Oxford Street
Cambridge, MA 02138, USA
Phone: 857-236-4652
Fax: 617-384-8859
ma.bind@mail.harvard.edu

**Running title:** Bridging observational studies and randomized experiments

**Acknowledgments:** Research reported in this publication was supported by the Ziff fund at the Harvard University Center for the Environment and by the Office of the Director, National Institutes of Health under Award Number DP5OD021412, NIH RO1-AI102710, and NSF IIS 1409177. The content is solely the responsibility of the authors and does not necessarily represent the official views of the National Institutes of Health. We also thank Profs. Speizer and Dockery for allowing us to use the data set and for providing useful comments.

**No conflicts of interest.**

**ABSTRACT**

The health effects of environmental exposures have been studied for decades, typically using standard regression models to assess exposure-outcome associations found in observational non-experimental data. We propose and illustrate a different approach to examine causal effects of environmental exposures on health outcomes from observational data. Our strategy attempts to structure the observational data to approximate data from a hypothetical, but realistic, randomized experiment.

This approach, based on insights from classical experimental design, involves four stages, and relies on modern computing to implement the effort in two of the four stages. More specifically, our strategy involves: 1) a conceptual stage that involves the precise formulation of the causal question in terms of a hypothetical randomized experiment where the exposure is assigned to units; 2) a design stage that attempts to reconstruct (or approximate) a randomized experiment before any outcome data are observed, 3) a statistical analysis comparing the outcomes of interest in the exposed and non-exposed units of the hypothetical randomized experiment, and 4) a summary stage providing conclusions about statistical evidence for the sizes of possible causal effects of the exposure on outcomes.

We illustrate our approach using an example examining the effect of parental smoking on children's lung function collected in families living in East Boston in the 1970s. Our approach could be credibly applied to less than 20% of the children in the families but found a realistic and important decrease in mean FEV-1 among children with parents who smoked vs. parents who did not smoke.

To complement the traditional purely model-based approaches, our strategy, which includes outcome free matched-sampling, provides workable tools to quantify possible detrimental exposure effects on human health outcomes especially because it also includes transparent diagnostics to assess the assumptions of the four-stage statistical approach being applied.

# 1. INTRODUCTION

Many studies have reported *associations* between environmental exposures and health outcomes using standard regression models analyzing non-randomized data, which is the norm in the field of environmental epidemiology because of ethical or logistic concerns about enforcing randomized assignment. However, *causal* relationships between environmental exposures and outcomes characterizing human health, although more difficult to estimate, are always the actual estimands in the field of environmental epidemiology, and moreover, estimates of these effects are expected by readers of epidemiological journals interested in policy implications. Here we consider an approach that explicitly attempts to estimate the causal effects of parental smoking on children's lung function, a causal question that is important, yet unanswered by extant analyses because past epidemiological studies have reported discordant estimates.[1] Providing accurate estimates of the causal effects of children's exposure to parental smoking is crucial to risk assessors. Although our analytic approach does not directly address the effects of specific interventions to curtail parental smoking, it does implicitly suggest, in the fourth stage, possible interventions to reduce the consequences on health outcomes. The causal versus associational nature of this relationship is reflected by the assertion that no matter what background characteristics lead to children's exposure to smoking parents, excess morbidity would be reduced if preventative interventions were implemented.

The general framework that we consider in this paper is sometimes called the "Rubin Causal Model"[2-5] for work done in the 1970's. This approach using potential outcomes to define causal effects was originally proposed by Neyman in 1923[6] but its use was restricted to randomized experiments until Rubin extended it to define causal effects in general.[7] To address causality, the key insight is to (multiply) impute the missing potential outcomes for each unit, i.e., what the outcome would have been under the other (meaning, not taken) treatment. In contrast, most published epidemiological studies model only the observed outcome data (i.e., not the potential outcomes) using associational models implicitly assuming that *"association implies some sort of causation"*. The main focus of this manuscript is to illustrate how to incorporate conceptual and design stages in observational studies prior to any analysis stage examining outcome data, which follows previous logic proposed by Rubin.[8, 9] Our approach transports established insights from classical experimental design, which revolutionized many empirical fields from 1925 to 1960.[10-13] Specifically, we use design strategies that were suggested in the late 1960s and early 1970s,[14-17] and compare the results to the results obtained by the standard strategy used in environmental epidemiology.

## 2. OUR SUGGESTED APPROACH – STEPS TOWARDS BIOLOGICAL EVIDENCE

Consider the specific environmental health example to estimate the causal effect of exposure to one causal factor, parental smoking, on children's lung function, assessed using forced expiratory volume in one second (FEV-1) in children using data collected in East Boston in the 1970s and previously analyzed decades ago[18] and more recently used for pedagogical purposes.[19] To our knowledge, all reports analyzing these data lacked both a conceptual stage and a design stage, and focused on the conclusions based on standard regressions generated from a simple analysis stage.

### 2.1. The standard analysis stage strategy

The standard epidemiological approach to such data has lung function as the outcome variable and has parental smoking and background variables as predictors in generalized linear or additive regression models. Association estimates are obtained, but:

*i) Are these estimated effects of similar magnitude to those that would be obtained if the researcher had conducted a real randomized experiment?*

*ii) What are the assumptions underlying standard regression models and are they straightforward or opaque?*

*iii) What are the precise meanings and robustness of the reported statistical summaries (e.g., p-values)?*

## 2.2. Overview of our approach leading to objective and valid causal inference under stated assumptions

In contrast, our approach proposes an analysis strategy with four transparent, distinct, and ordered stages, following implicit advice in classical texts on experimental design (e.g., Fisher,[10,11] Kempthorne,[12] Cochran and Cox,[13] Box et al.[20]) and a more recent text extending this perspective to non-randomized studies.[5]

> 1) A *conceptual* stage that involves the precise formulation of the causal question (and related assumptions) using potential outcomes and described in terms of a hypothetical randomized experiment where the exposure is randomly assigned to units; this description includes the timing of random assignment and defines the target population; no computation is needed at this stage.
>
> 2) A *design* stage that attempts to reconstruct (or approximate) the design of a randomized experiment before any outcome data are observed (that is, with unconfounded assignment of exposure using the observed background and treatment assignment data); typically, heavy use of computing is needed at this stage, e.g., for multivariate matched sampling and extensive balance diagnostics.
>
> 3) A *statistical analysis* stage defined in a protocol explicated before seeing any outcome data, comparing the outcomes of interest in similar (e.g., hypothetically randomly divided) exposed and non-exposed units of the hypothetical randomized experiment; this stage is the one that most closely parallels the standard model-based analyses but uses more flexible methods.
>
> 4) A *summary* stage providing conclusions about statistical evidence for the sizes of possible causal effects of the exposure; no computing is required at this stage, just thoughtful summarization, e.g., focusing on what actual world interventions could be implemented to curtail any untoward causal effects of the exposure.

## 3. OUR APPLICATION: THE EFFECT OF PARENTAL SMOKING ON CHILDREN'S LUNG FUNCTION

Our data set comprises 654 children and young adults, 318 females and 336 males, with 10% having parents who smoke. The children's ages range between 3 and 19 years old. Regarding the heights of the children, the mean is 61 inches and they range between 46 and 74 inches.

### 3.1. Overview of the four stages in our example

*3.1.1. Conceptual stage: precise formulation of the causal question*

Several hypothetical randomized experiments where "enforced smoking cessation" is randomly assigned to parents, can be conceptualized (e.g., Bernouilli trial, completely randomized experiment, stratified randomized experiment, paired randomized experiment). At this initial stage, the <u>plausibility</u> of the reconstructed hypothetical randomization is important because we want to convince the reader of that position on which the entire analysis is predicated. Different timings of the random assignment can be imagined (e.g., before or after conception of the child) and different target populations from which the sample of 654 children was drawn can be considered.

*3.1.2. Design phase: reconstruction of the hypothetical experiment*

To address causality, our position is that we need to start by approximating the ideal conditions of a randomized experiment, which demands unconfounded assignment of exposure given observed covariates. Unconfoundedness of the exposure's assignment can be achieved approximately using matching techniques aiming to create exchangeable groups (e.g., strata, pairs) of exposed (to parental smoking) and non-exposed units with randomly different values of pre-exposure (background) covariates; in our simplified example, such covariates include age, height, and sex. That is, we attempt to create exchangeable exposed and unexposed groups or matched pairs of children, one member (or part) of each group or pair is randomly assigned to smoking parents and the other member to non-smoking parents but matched on age, height, and sex. Some earlier methods and associated theory are summarized by Rubin[17] and Rosenbaum,[21] and more recent approaches are given in Sekhon[22] and Hansen et al.[23]. If the matching strategy creating two such *identical* groups or pairs of children is entirely successful, there can be no confounding with respect to the background variables that we used for matching. It is obviously not ethical to transfer children to different parents, but perhaps it is plausible that non-smoking characteristics of smoking and non-smoking parents have no effect on children's lung function. At least we should be explicit about such important, but typically implicit, assumptions.

*3.1.3. Analysis phase*

We start by using computationally flexible techniques, such as statistical matching, to achieve balanced distributions of the background variables in the exposed and unexposed children. The most straightforward analysis examines the difference in lung function between the exposed children and the unexposed children, and these are then averaged over all children to obtain an estimate of the average causal effect. Randomization-based inference can be conducted using modern computing techniques[5] to test the sharp null hypothesis that exposure to parental smoking has absolutely no effect, relative to no smoking exposure, on children's lung function. Frequentist or Bayesian regression models can also be used at that stage in order to increase efficiency, by removing residual confounding that was not adequately addressed during the design stage, e.g., allowing treated and controls to have separate regression slopes and separate residual variances.[15-17, 24, 25] It is critical that the analysis stage needs to be specified in a protocol explicated before seeing any outcome data.

*3.1.4. Causal conclusion*

If one observes a significant difference in average lung function outcome between these exchangeable groups or matched pairs (i.e., a difference that would be a rare event in the hypothetical randomized experiment if there were no effect of exposure), it is natural to attribute that difference to the differential exposures to parental smoking, and critically, to propose that the negative effect could be ameliorated by the introduction of some hypothetical intervention to curtail smoking, yet to be debated.

3.2. Details of the three first stages in our example

*3.2.1. Six hypothetical experiments (first stage)*

Various possible interventions to curtail parental smoking are now discussed. An important issue related to the timing of the observational data collection arises in this setting because children's characteristics (such as *age*, *height*, and *sex*) in our data set are actually known only *a posteriori*, that is, after assignment to the exposure. If we assess whether children are similar with respect to variables measured *after* the assignment of exposure, we need to assume, for the validity of simple analyses, that these variables are not affected by the exposure. Note that although this assumption may not be plausible for the exposure of parental smoking and the variable height, we found no evidence of parental smoking influencing height in our data set after applying our suggested approach but considering *height* as an outcome with *age* and *sex* as covariates.

Hypothetical experiment A: One *hypothetical* completely randomized experiment (with $N_{Smoking}=65$ children with smoking parents and $N_{Non-Smoking}=589$ children with non-smoking parents) that we can imagine involves intervening on smoking households before they have children and randomizing them to stop smoking with probability 9/10, and thus with probability 1/10 to continue to smoke.

We can see that formulating a hypothetical intervention can be challenging. First, it should be plausible enough to convince readers to continue reading. However, we believe it is one the most interesting and scientifically, not mathematically, relevant steps for epidemiological researchers. Note that whatever hypothetical intervention you posit for the experiment underlying your dataset, you are assuming that you will obtain essentially the same analytic answer for all versions of that hypothetical experiment. That is, there is a hidden assumption at this stage that, whichever version of the hypothetical intervention you choose, it will lead to approximately the same estimated causal effect. More precisely, in our example, can you argue that the hypothetical intervention assuming that the population consisted of only smoking parents who were assigned to stop smoking with probability 9/10 (and they all complied) would lead to the same conclusion as if the population consisted of only non-smoking parents who were assigned to smoke with probability 1/10 with full compliance? The latter would be clearly unethical considering what we now know about smoking exposure. But this question emphasizes the type of question you should be willing to entertain and answer. Actually, we do not consider Hypothetical Experiment A plausible. For reason discussed shortly, perhaps discarding unexposed children with background characteristics that are unlike the exposed children and vice versa would improve the plausibility of a hypothetical experiment?

Hypothetical experiment B: Another *hypothetical* completely randomized experiment could have resulted in exposed children with background covariates that are within the range of the background covariates of the unexposed children and unexposed children with background covariates that are within the range of the background covariates of the exposed children. That is, suppose we selected boundaries for the covariates *age* and *height*, and restricted the 361 children to fall within those boundaries. This strategy led to $N_{Smoking}$=61 children with smoking parents and $N_{Non-Smoking}$=300 children with non-smoking parents. At this point, an underlying hypothetical experiment that generated the data was not yet considered plausible; the specific reasons will be explained in section 3.2.4.

Hypothetical experiment C: Another *hypothetical* randomized experiment could have resulted in non-smoking parents with background covariates that are within certain strata defined by the background covariates of the smoking parents. This formulation is described more precisely in section 3.2.2, part b), and led to $N_{Smoking}$=57 children with smoking parents and $N_{Non-Smoking}$=216 children with non-smoking parents.

Other hypothetical randomized experiments would also intervene on smoking parents before their child's conception; we describe two such experiments. First, Hypothetical experiment D.1, a completely randomized experiment with balanced groups (e.g., creating two equal-sized groups of parents similar on background characteristics, that is, $N_{Smoking}=N_{Non-Smoking}$=63 children). Or second, Hypothetical experiment D.2: a rerandomized experiment with two equal-sized groups of similar parents (with $N_{Smoking}=N_{Non-Smoking}$=63) for which the randomized allocations are allowed only when parents' covariates (e.g., height) mean differences between smokers and non-smokers are within some *a priori* defined calipers.

Another *hypothetical* randomized experiment, Hypothetical experiment E, would intervene after the child's conception, from the point in time for which we know the child's gender, and would have a paired-randomized experiment where a coin flip determines which parents of a pair of two similar parents expecting a child with same gender is exposed to still-smoking parents, with $N_{Smoking}=N_{Non-Smoking}$=63 children).

We define the "finite population" as the population being randomized in each of the reconstructed hypothetical experiments. The super-population is a hypothetical "infinite population" from which the finite population is drawn.

*3.2.2. Several different design phase strategies (second stage)*

a) No design stage (a)

The standard approach in environmental epidemiological lacks both a conceptual stage and a design stage and simply focuses on associations gleaned from observed data ($N_{children}$=654).

b) Trimming (b)

A relatively naïve strategy attempts to eliminate units from one group (i.e., treated or control) outside the range of the other group with respect to background covariates by trimming "outlier" units. For example, in our data set, although the ages of girls with non-smoking parents range from 3 to 18 years old, the ages of girls with smoking parents range from 10 to 19 years old; the ages of boys with non-smoking parents range from 3 to 19 years old, whereas the ages of boys with smoking parents range from 9 to 18 years old. Similarly, the heights of girls with non-smoking parents range from 46 to 71 inches, whereas the heights of girls with smoking parents range from 60 to 69 inches; the heights of boys with non-smoking parents range from 47 to 74 inches, whereas the heights of boys with smoking parents range from 58 to 72 inches. Therefore, to restrict imbalance with respect to age and height in the exposed vs. non-exposed groups, we included girls with ages between 10 and 18 years and heights between 60 and 69 inches, and included boys with ages between 9 and 18 years and heights between 58 and 72 inches; these restrictions leave us with 361 units out of 654. Note that, at that stage, any remaining imbalance in any background variable (e.g., age) between the exposed and non-exposed groups still limits our ability to assert that the "hypothetically randomized" exposure was the sole reason for the lack of balanced background covariates between children with smoking parents and children with non-smoking parents.

c) Stratified matching (c)

Another approach is to go beyond trimming and construct discretized covariates and thus strata in which these discretized background covariates are balanced. This strategy, essentially proposed decades ago in the context of missing data as "hot deck" imputation[26] and then for matching in causal inference by Cochran,[14] has recently become popularized and renamed with the oxymoron "coarsened exact matching".[27] This approach eliminated 381 children out of 654.

d) Propensity score one-to-one matching after overlap assessment and discarding (d)

A one-to-one matching strategy with calipers[28] on the estimated propensity score,[21] for instance estimated by a logistic regression that regresses parental smoking on the available covariates in the dataset (e.g., *age*, *height*, *sex,* and non-linear functions of them), but no outcome variables, can also be used in the design stage. A more parsimonious (and therefore simpler to interpret) model including *age*, *height*, and *sex* rather than *age*, *age²*, *height*, *height²*, *sex*, *sex\*age*, and *sex\*height* was favored by us based on likelihood ratio tests, as suggested in Imbens and Rubin.[5] We removed 156 "outlier" children (i.e., 154 with non-smoking parents and two with smoking parents) with estimated propensity scores that did not overlap with the other group (see Figure 1 showing the estimated propensity score distributions among the children with smoking parents and non-smoking parents before and after removing the "outlier" children). We required covariates balance within a caliper equal to one standard deviation of the raw propensity score. The approach led to 63 exposed children and 63 unexposed children with similar background characteristics at the group level, not necessarily pair by pair, even though pairs were used to construct overlapping treatment and control groups.

e) Optimal pair matching after overlap assessment and discarding (e)

After removing "outlier" children, another matching strategy creates "optimal" pairs of children, where optimal means minimizing the squared Mahalanobis distances between paired exposed and unexposed children with respect to the covariates *age*, *height*, and *sex*.[23] The "optimal" pairing matched 63 exposed children to 63 similar unexposed children. This approach may have the advantages of directly creating well-matched pairs with an *a priori* optimization criteria (e.g., squared Mahalanobis distance), or equivalently removing pairs not satisfying this criterion; thereby having some flavor of the rerandomization approach[29].

*3.2.3. Description of the final resulting datasets across hypothetical experiments / design stage methods*

A summary of the characteristics of the units arising from each hypothetical experiment resulting from each design stage method is presented in Table 1. When trimming the outlying units, the dataset is reduced from 654 to 361 children (i.e., 55% of the children remain) with an increased mean age, mean height, ratio of boys to girls, and ratio of smoking parents to non-smoking parents. The stratified matching strategy reduced the dataset further to 273 children with characteristics similar to the trimmed dataset. The propensity score and optimal pair matching approaches reduced the dataset even more to 63 pairs of children (i.e., 126 total children, 20% of the original children) with similar age and height characteristics as in the trimmed dataset but with fewer children with non-smoking parents and fewer boys.

*3.2.4. Initial assessment of plausibility of the reconstructed hypothetical randomized experiments*

To start the assessment of the plausibility of each hypothesized experiment, we examine whether the background covariates are successfully balanced between treatments. For each hypothetical experiment, we present the mean and standard deviation of age, height, and of female-male proportion in the exposed and unexposed groups (Table 2).

The reconstructed hypothetical randomized experiment (A) is <u>not plausible</u> for our data because the East Boston study population did not consist of parents all of whom smoked at one time. The background characteristics of the study population in the original data set are also inconsistent with a "good" randomization because, for instance, children with smoking parents are significantly older, and thus, not surprisingly, taller than children with non-smoking parents (first row of Table 2). The reconstructed hypothetical randomized experiment (B) is also <u>not plausible</u> because the background characteristics of the study population in the trimmed data set is inconsistent with a "good" randomization; children with smoking parents are still significantly older, taller than children with non-smoking parents (second row of Table 2). The reconstructed hypothetical randomized experiment (C) is also <u>not plausible</u> because the background characteristics of the study population of smoking and non-smoking parents in the described experiment is inconsistent with a "good" randomization; children with smoking parents are still significantly older and taller than children with non-smoking parents (third row of Table 2). The last three reconstructed hypothetical randomized experiments (e.g., D.1, D.2, and E) could be <u>plausible</u> because the background characteristics of the study population of smoking and non-smoking parents in the described experiments are consistent with fairly "good" randomizations; children with smoking parents are not significantly different from children with non-smoking parents (fourth and fifth rows of Table 2).

For each reconstructed hypothetical experiment, plausible or not, we compare the estimated averaged causal effects (ACEs) using standard analysis strategies. However, for illustrating the Fisherian and Bayesian inferences, for reasons of conciseness, we chose to focus on the three plausible reconstructed randomized experiments, that is, we consider only the matched-sampling datasets obtained via the propensity score matching (d) (corresponding to hypothetical completely randomized experiment (D.1) and rerandomized experiment (D.2)), and the optimal pair matching (e) (corresponding to hypothetical paired-randomized experiment (E)) approaches.

*3.2.5. Additional assessment of balance in covariates*

Many methods have been proposed to assess balance in covariates (some reviewed by Imbens and Rubin[5]). We also calculated the standardized mean differences between the exposed and unexposed children (before and after matching on *age*, *height*, and *sex* using propensity score calipers and optimal pairing) of the variables *age*, *age$^2$*, *height*, *height$^2$*, *sex*, *sex\*age*, and *sex\*height*. The top panel of Figure 2 shows that the standardized mean differences between exposed and non-exposed children were reduced after propensity score matching for all variables included when estimating the propensity score (i.e., *age*, *height*, and *sex*), as well as for the variables not included in the propensity score (i.e., *age$^2$*, *height$^2$*, *sex\*age*, and *sex\*height*) because these were correlated with the estimated propensity score. Note that smaller calipers could have been chosen but minimal improvement was achieved with respect to overall covariate balance. The "Love" plot for the optimal matching strategy (bottom panel of Figure 2) suggests excellent balance between the exposed and unexposed children.

Another way of assessing balance for continuous covariates, which can provide more detailed insights than the standard "Love" plots presented in Figure 2, is to present the empirical distributions of *age* and *height* for the exposed vs. non-exposed children before and after matching (Figures 3 and 4). For conciseness, we presented these distributions of the continuous variables *age* and *height* among the exposed and unexposed children before and after matching on *age*, *height*, and *sex* only for experiments 4 and 5 (i.e., for the propensity score caliper (d) and optimal pairing (e) approaches). We also reported Kolmogorov-Smirnov tests to assess whether the univariate distributions of the variables for the exposed children differ from those distributions for the unexposed children (before and after matching using propensity score caliper (d) and optimal pairing (e)). As shown in Figure 3, although the Kolmogorov-Smirnov test comparing the *age* distributions for the non-exposed vs. exposed children in the original data set was highly significant at traditional levels, it was not so after matching using the propensity score and after constructing optimal pairs. Similarly, as shown in Figure 4, the Kolmogorov-Smirnov test comparing the *height* distribution for the non-exposed vs. exposed children in the original data set was highly significant, but it was not after propensity score matching or after optimal pair matching. The distributions of the squared Mahalanobis distances between propensity score (top panel) vs. optimal pairs (bottom panel) are presented in Figure 5. Although the range of pairwise squared Mahalanobis distances is between 0 and 12 for the propensity score matched pairs, with the optimal pair matching approach, the range of these squared distances is between 0 and 2, which suggests better pairing.

*3.2.6. Analysis phase (third stage): various standard regression-based outcome analysis-phase strategies at the super-population level*

<u>i) T-test / crude regression analysis (i.e., no covariate adjustment)</u>

An initial t-test can be conducted comparing the mean FEV-1 among children with smoking parents to the mean FEV-1 among children with non-smoking parents. This is equivalent, assuming that the treatment effect is constant and additive for all units and that the residual variances in both groups are the same, to regressing the dependent variable, FEV-1, on the indicator for exposure of interest "parental smoking", and examining the size and statistical significance of the coefficient of the indicator.

ii) Standard linear regression model with simple linear adjustment

The second analysis regresses the dependent variable FEV-1 on the exposure of interest "parental smoking" as in analysis (i) but also linearly "adjusts" for the three covariates available in the dataset, i.e., age, height, and sex, by including them in the regression model, and making the analogous assumptions as with the first analysis. The distributions of the outcome of interest FEV-1 across children with parents who smoke and not, stratified by *sex*, are presented in the Supplementary Figure 1. We also assessed the significance of interaction terms between parental smoking and the three covariates, and found limited evidence of interactions ($p_{interaction}=0.14$ for *smoking\*age*, $p_{interaction}=0.10$ for *smoking\*height*, and $p_{interaction}=0.26$ for *smoking\*sex*). We also found little evidence against the linearity assumption of the associations between 1) *age* and FEV-1, and 2) *height* and FEV-1 (see Supplementary Figure 2). Other versions of this regression were investigated in the original dataset, that is, using all 654 units (i.e., omitting conceptual and design stages).[19]

*3.2.7. Analysis-phase strategies (third stage) at the finite-population level*

i) Analysis using Fisherian (Fiducial) inference in the finite population

Because there were three plausible hypothetical randomized experiments, we perform randomization-based tests assuming the data arise from: i) the complete randomization experiment (D.1), ii) the rerandomized experiment (D.2), and iii) the pairwise randomized experiment (E). That is, we test the Fisher null hypothesis of no effect of parental smoking on children's FEV-1 in the finite population sample by performing a stochastic proof by contradiction. We first assume the null hypothesis of absolutely no effect of treatment versus control, so that we know all potential outcomes and thus know what the value of any test statistic would be obtained under any treatment assignment. Then, for the completely randomized experiment (D.1), we permuted the treatment assignment among the 126 children such that half of them get exposed and obtain 126-choose-63 different treatment assignments. Similarly, for the hypothetical rerandomized experiment (D.2), we rerandomized the 126 children such that half of them get exposed but the two groups have similar background covariates' means. Finally, for the paired randomized experiment, we choose one member of each of the 63 pairs to be considered treated, and thereby obtain $2^{63}$ different treatment assignments. We conducted 10,000 random draws of permuted 1) completely randomized (D.1), 2) rerandomized (D.2), and 3) pair randomized treatment assignments (E), and calculate the following statistic in each permuted allocation:

1) $T_{\text{t-completely randomized D.1}}$ = t-test statistic comparing the mean FEV-1 among exposed and unexposed children (different group variances),

2) $T_{\text{t-rerandomized D.2}}$ = t-statistic of the regression coefficient of *smoking* when regressing FEV-1 on smoking, age, height, and sex, and

3) $T_{\text{t-paired randomized E}}$ = paired t-test statistic comparing the means FEV-1 among exposed vs. unexposed children.

We obtain Fiducial intervals by inverting the sharp null hypothesis tests for different constant additive effects as described in Imbens and Rubin.[5]

ii) Analysis using Bayesian inference to estimate the posterior distribution of the average causal effect (ACE) and its 95% probability interval in the finite population

We now consider the Bayesian approach initially proposed by Rubin[30] and described in Imbens and Rubin.[5] Briefly, we first specify distributions for the potential outcomes conditional on covariates, here for simplicity independent and identically distributed normal ones. Because we consider only the plausible hypothetical randomized experiments (D.1, D.2, and E) in this section, we assume ignorable exposure assignment (i.e., P(Smoking$_i$=1 | FEV-1$_i^{obs}$, FEV-1$_i^{mis}$, Age$_i^{obs}$, Height$_i^{obs}$, Sex$_i^{obs}$) = P(Smoking$_i$=1 | FEV-1$_i^{obs}$, Age$_i^{obs}$, Height$_i^{obs}$, Sex$_i^{obs}$), where FEV-1$_i^{obs}$ and FEV-1$_i^{mis}$ represent the observed and missing FEV-1 potential outcomes for the $i^{th}$ unit).[30] We impute the missing potential outcomes among the exposed and non-exposed groups separately, allowing for different normal models (conditional on the intercept and the three covariates available in the dataset, i.e., *age*, *height*, and *sex*), that is, different means ($\mu_{i,Smoking} = \beta_{Smoking} X_i$ and $\mu_{i,Non-smoking} = \beta_{Non-Smoking} X_i$, where $X_i$ represents the constant, *age*$_i$, *height*$_i$, and *sex*$_i$) and different variances in the exposure groups ($\sigma_{Smoking}^2$ and $\sigma_{Non-smoking}^2$). The goal is to draw multiple values of FEV-1$_i^{mis}$ conditional on FEV-1$_i^{obs}$, Smoking$_i^{obs}$, Age$_i^{obs}$, Height$_i^{obs}$, Sex$_i^{obs}$, and the parameters $\beta_{Smoking}$, $\beta_{Non-Smoking}$, $\sigma_{Smoking}^2$, $\sigma_{Non-smoking}^2$. To accomplish this, we need to calculate the posterior distribution for the parameters. We assume flat priors for the parameters $\beta$ and $\sigma^2$, that is, $p(\beta_{Smoking}, \sigma_{Smoking}^2) \propto \sigma_{Smoking}^{-2}$ and $p(\beta_{Non-Smoking}, \sigma_{Non-Smoking}^2) \propto \sigma_{Non-Smoking}^{-2}$. We use two separate Gibbs samplers to impute: 1) the missing control potential outcomes among the treated, and 2) the missing treated potential outcomes among the controls, reflecting independent prior distributions for these parameters.

For instance, to impute the <u>control</u> missing potential outcomes, that is, FEV$_i$-1$^{mis}$ = FEV$_i$-1[Smoking$_i$=0)] among the exposed children, 1) we draw $\sigma_{Non-smoking}^2$ such that $1/\sigma_{Non-smoking}^2 \sim \{1/[(n_{Non-Smoking} - 4) s_{Non-Smoking}^2]\} \chi^2$ with $n_{Non-Smoking} - 4$ degrees of freedom, where $n_{Non-Smoking}$, $s_{Non-Smoking}^2$ are the number of children with non-smoking parents and the FEV-1 sample variance among the children with non-smoking parents, respectively; 2) we then draw $\beta_{Non-Smoking}$ conditional on $\sigma_{Non-smoking}^2$, FEV$_i$-1$^{obs}$, Smoking$_i^{obs}$, $X_i^{obs}$ from a normal distribution with mean equal to $[(X_{Non-Smoking}^T X_{Non-Smoking})^{-1} X_{Non-Smoking}^T FEV_i\text{-}1_{Non-Smoking}]$ and variance-covariance matrix $[X_{Non-Smoking}^T X_{Non-Smoking})^{-1} \sigma_{Non-smoking}^2]$, and finally, 3) draw the missing control potential outcomes among the treated; that is, for unit *i* such Smoking$_i$=1, we draw FEV-1$_i^{mis}$ conditional on FEV-1$_i^{obs}$, W$_i$, β$_{Non\text{-}Smoking}$, and σ$_{Non\text{-}smoking}^2$ independently from a normal distribution with mean [X$_i^{obs}$ β$_{Non\text{-}Smoking}$] and variance σ$_{Non\text{-}smoking}^2$.

At each replication, we impute the missing potential outcomes in <u>both groups</u> and calculate the average causal effect (ACE), i.e., the mean difference in FEV-1 among all children when having smoking parents vs. when having non-smoking parents. We repeat this procedure 10,000 times and thereby obtain 10,000 draws of the ACE.

<u>iii) Mixing the Bayesian and Fisherian approaches</u>

The Bayesian approach relies on the model specification to be approximately correct, whereas the Fisherian procedure provides a non-parametric procedure to test the sharp null hypothesis. We propose to use a different, and possibly more interesting, statistic than T$_{t\text{-completely randomized D.1}}$, T$_{t\text{-rerandomized D.2}}$, and T$_{t\text{-paired randomized E}}$ calculated from the approximated Bayesian posterior distribution of the average causal effect to test the sharp null hypothesis, T$_{t\text{-Bayesian}}$ = | posterior mean of the ACE | / standard deviation of the ACE. The idea to use a statistic based on a model for the Fisher test goes back at least to Brillinger, Jones and Tukey.[31]

**3.3. Results from our example**

*3.3.1. Estimated average causal effects (ACE) and associated asymptotic 95% confidence intervals in the super-population (see Table 3)*

<u>a) No design stage</u>

The first two rows of Table 3 summarize the two analyses with no design stage, and both indicate a beneficial or uncertain effect of smoking parents on children's FEV-1.

<u>b) Trimming</u>

From Table 3, the trimming approach provides estimated ACEs that indicate essentially slightly beneficial or uncertain effects of parental smoking on children's FEV-1.

c) Stratified matching

From the fifth and sixth rows of Table 3, we see that, with the stratified matching strategy, the estimated ACEs indicate some possible negative effects of parental smoking on children's FEV-1.

d) Propensity score matching

With 126 units, but restricting the data to pairs of children who are "similar" with respect to age, height, and sex, the propensity matched sampling approach estimates the crude and adjusted estimated effects of parental smoking on children's FEV-1 to be negative. That is, the mean FEV-1 among children with parents who smoke was estimated to be lower than the mean FEV-1 among children with non-smoking parents. The squared Mahalanobis distances between propensity score pairs are greater for the negative estimated paired causal effects as shown in Figure 6, suggesting some "outlying" pairs.

e) Optimal pair matching

With 63 "optimal" pairs, the crude and adjusted estimated effects of parental smoking on children's FEV-1 also suggest negative effects.

*3.3.2. Fisherian and Bayesian inferences in the finite population*

i) Fisherian (Fiducial) inference in the finite population

The approximated null randomization distributions of the chosen statistics $T_{\text{t-completely randomized D.1}}$, $T_{\text{t-rerandomized D.2}}$, and $T_{\text{t-paired randomized E}}$ (based on 10,000 draws of the permuted treatment assignment) are presented in Figure 7. The proportion of the equiprobable treatment allocations under randomized assignment that led to values of the statistics, $T_{\text{t-completely randomized D.1}}$, $T_{\text{t-}}$ rerandomized D.2, and $T_{\text{t-paired randomized E}}$, as large or larger than the observed statistic $T^{\text{obs}}_{\text{t-completely randomized D.1}}=1.57$, $T^{\text{obs}}_{\text{t-rerandomized D.2}}=1.66$, and $T^{\text{obs}}_{\text{t-paired randomized E}}=2.12$ were equal to p-value$_{\text{completely randomized D.1}}=0.12$, p-value$_{\text{rerandomized D.2}}=0.10$, and p-value$_{\text{paired randomized E}}=0.04$, respectively, all suggesting significant effects of parental smoking.

Inverting these sharp null hypothesis tests for different values of average causal effects across the three reconstructed randomized experiments led to 95% Fiducial intervals equal to [-0.52 to 0.06]$_{\text{completely randomized D.1}}$, [-0.33 to 0.03]$_{\text{rerandomized D.2}}$, and [-0.37 to -0.02]$_{\text{paired randomized E}}$, again suggesting negative effects of parental smoking on children's FEV-1.

ii) Bayesian inference for the posterior distribution of the average causal effect (ACE) and its 95% probability interval in the finite population

The posterior distributions of the average causal effect (ACE) using the matched-sampling datasets obtained via the propensity score (top panel) and the optimal pair matching (bottom panel) approaches are presented in Figure 8. The posterior means are -0.16 and -0.18 and the 95% probability intervals are [-0.29 to -0.04] and [-0.30 to -0.06], respectively, suggesting fairly clear evidence of negative effects of parental smoking on children's FEV-1.

iii) Mixing the Bayesian and Fisherian approaches

The approximated null randomization distributions of the chosen statistics $T_{\text{t-completely randomized D.1 and Bayesian}}$, $T_{\text{t-rerandomized D.2 and Bayesian}}$, and $T_{\text{t-paired randomized E and Bayesian}}$ (based on 10,000 draws of the permuted treatment assignment) are presented in Figure 9, respectively. The proportion of the equiprobable treatment allocations under randomized assignment that led to statistics $T_{\text{t-completely randomized D.1 and Bayesian}}$, $T_{\text{t-rerandomized D.2 and Bayesian}}$, and $T_{\text{t-paired randomized E and Bayesian}}$ with as large or larger values than the observed statistic $T^{\text{obs}}_{\text{t-completely randomized D.1 and Bayesian}} =2.39$, $T^{\text{obs}}_{\text{t-rerandomized D.2 and Bayesian}} =2.31$, and $T^{\text{obs}}_{\text{t-paired randomized E and Bayesian}} =2.84$ were equal to p-value$_{\text{completely randomized D.1}}$ and Bayesian $=0.09$, p-value$_{\text{rerandomized D.2 and Bayesian}}=0.10$, and p-value$_{\text{paired randomized E and Bayesian}}=0.04$, respectively, again suggesting parental smoking is not good for children's FEV-1.

## 4. DISCUSSION

Even though our approach uses fewer units in the analysis phase (i.e., third stage) compared to the standard model-based approach without conceptual or design phases, it can still reach relevant conclusions, arguably more credible than the standard ones. Our results contrast with the naive idea that more units of analysis always bring more statistical power to detect causal effects. Our final causal conclusion appears to support the reported associational estimate in the well-known Harvard Six Cities longitudinal study,[32] in which Wang et al. reported that each pack per day smoked by the mother was associated with a reduction of 0.4% [95%CI: -0.9% to 0.1%] in FEV-1 among children six to ten years old (after "adjusting" for age, height, city of residence, and parental education).

Once causality is suspected, the next step is to acquire medical knowledge, for instance, trying to understand biological mechanisms explaining why exposure to parental smoking causes reduced lung function (e.g., via smoking-specific inflammatory biomarkers). Also, interventions that may curtail smoking can be explored, for instance by trying to predict the occurrence of smoking among parents using the background covariates to predict smoking.

Our approach with conceptual and design phases facilitates an approximation to the ideal conditions of a randomized experiment and has the tremendous advantages that these phases can be conducted blind to the outcome data and that their formulation relies on creative thinking by the environmental epidemiologist. Obviously, inferences are restricted to children who remain in the sample. Extrapolation to children with covariate values beyond values observed in the matching children should generally be done with great caution because the data do not provide direct information for treated children without control matches. This is one advantage of classical randomization-based inference advocated here vs. the more common purely model-based approaches using the entire data set. Fisher randomization-based p-values associated with explicit designs can be easily conceptualized and obtained, and no asymptotic distributional assumptions are used. In our approach, as in the design of randomized experiments, we eschew the use of outcome variables to create the matched pairs.[33] Instead, we attempt to recreate hypothetical completely randomized, rerandomized, and matched pair randomized experiments. This process was implied more than a half century ago by Dorn's 1952 sage advice, repeated by Cochran,[34] "*How would the study be conducted if it were possible to do it by controlled experimentation?*".

A causal investigation needs to examine the implicit assumption that the hypothetical set of control children is effectively stochastically identical to the set of exposed children on all their observed background variables. This assumption is explicit, transparent, and readily assessed by simple visual diagnostics. For instance, Figure 2 shows the effect of matching on the standardized mean differences between exposed and non-exposed children for the covariates *age*, *height*, and *sex* (allowing for linear and quadratic relationships, as well as interactions). If all covariates and their non-linear terms were as well matched, then a logical, although tentative, conclusion can be reached concerning the evidence that parental smoking was the cause of any discrepancies between the exposed and non-exposed children in lung function, in the sense that if we could eliminate parental smoking without any untoward consequences of the intervention, this difference in lung function would be found for experimental data. Figures 3 and 4 present the effect of matching (via propensity score and optimal pairing) on the distributions of the continuous covariates *age* and *height*, respectively, i.e., in this case, matching created almost identical *age* and *height* distributions for exposed and non-exposed children, which is ideal for eliminating any confounding arising from *age* and *height*.

We considered different methods using either stratification, propensity score caliper matching, or optimal pairing using Mahalanobis distance. In our data set, the optimal pairing led to very well-matched children and appears to be *ideal* for our data as a design stage procedure preceding the (multiple) imputation of the missing potential outcomes. In settings with more than three background covariates, minimizing the squared Mahalanobis distance will not be as satisfactory as in settings with low-dimension covariates because every unit is likely to be far apart on this full-rank metric,[17] so it may be better to minimize this distance within pairs in the same propensity score caliper only with respect to the continuous covariates (e.g., using the procedure proposed in Rosenbaum and Rubin[21]). Other balancing criteria could be used that may be more *relevant* to optimize than some function involving Mahalanobis distance. This optimized criterion-based rejection (OCBR) approach discarding units that do not satisfy the criterion may be attractive and flexible with respect to the choice of criterion because it can combine several criteria measuring covariates' balance. If the *a priori* optimization criteria would have combined diagnostics of covariates imbalance, such as 1) differences in covariates means and variances between exposed and unexposed, followed by, 2) Kolmogorov-Smirnov distances between continuous covariates distributions in the exposed vs. unexposed), these balancing diagnostics would automatically be satisfied by the procedure.

Some drawbacks of the OCBR strategy are that the approach is computationally intensive and currently lacks software implementation for *exotic* criterion. The optimal matching strategy also selects only one matched dataset, the one with minimum total squared Mahalanobis distance, which may restrict pure randomization-based inference. Future work should consider criterion-based rejection (CBR) approaches constructing matched datasets satisfying a balancing criterion instead of an optimization function.

Unmatched data from exposed children that have background characteristics that differ markedly from the background characteristics of unexposed children are discarded in our approach; yet such children values are automatically included in standard model-based regression, and their inclusion can distort the prediction of missing potential outcomes and therefore the causal conclusion. Also, even if the point and interval estimates were to agree numerically between our analysis and a standard analysis, the "results and associated conclusions" are not necessarily the same. Not only are our conclusions explicitly limited to children represented by groups or pairs that are well-matched, but the assumptions underlying the hypothetical randomized experiments are entirely transparent and accessible, as exemplified by Figures 2 to 6 and Table 2, and therefore facilitate discussions among scientists about their veracity.

We feel that our matched-sampling strategy, based on the hypothetical randomization that created the sets of exposed versus non-exposed units, followed by the analysis of data by randomization tests, relies on powerful and modern computing to implement both a) the creation and analysis of exchangeable groups or pairs, and b) the fiducial tests themselves. Of particular interest, these types of analyses using 1) matched-sampling techniques, 2) constructing a t-statistic summarizing the Bayesian analysis, and 3) performing non-parametric Fisherian inference, have apparently not been previously done, or even contemplated, in environmental epidemiology. Combining the Bayesian and Fisherian inference frameworks could lead better statistical properties.[35]

Our approach may have the potential to have a broad impact on the field of environmental epidemiology, because extensions implicitly propose a universal framework using classical ideas from randomized experiments to tackle causal questions examining the joint health effects of multi-factorial exposures (e.g., mixtures of indoor and outdoor air pollutants, weather conditions, physical activity, etc.). Here, when facing such questions, we propose embedding an observational data set within the context of a hypothetical multi-factorial randomized experiment. It is important to emphasize that this proposed approach is not restricted to relatively simple settings, but it generalizes to situations involving complex data structures (e.g., longitudinal data; "mediators" - to examine putative causal pathways; and high-dimensional data - to help discover the etiology of complex diseases or disorders).

## 5. CONCLUSIONS

We propose a logically and practically transparent, yet mathematically precise and rigorous, approach to study the health effects of the *multi-factorial environmental "exposome"* resulting in causal inferences that are valid under explicitly stated assumptions. This framework can be used to study biological mechanisms and susceptibility to complex diseases resulting from the joint effects of multiple environmental factors. Because of its conceptual links to hypothetical interventions, it can suggest policies for reducing environmental pollutants and thereby preventing diseases. Also because of its logical transparency, it should promote education across, and communication between, researchers and policy-makers.

1   **TABLES**

2   **Table 1: Description of the variables in the data sets across design stage methods**

| Variables used in each hypothetical experiment / design | Number of children | Min | 25th quantile | Mean | Median | 75th quantile | Max |
|---|---|---|---|---|---|---|---|
| **HYPOTHETICAL EXPERIMENT (A) / NO DESIGN (a)** | | | | | | | |
| Age (years) | 654 | 3 | 8 | 10 | 10 | 12 | 19 |
| Height (inches) | 654 | 46 | 57 | 61 | 62 | 66 | 74 |
| Parental smoking (0: no, 1: yes) | 654 | 0 | 0 | 10% | 0 | 0 | 1 |
| Male children (0: no, 1: yes) | 654 | 0 | 0 | 51% | 1 | 1 | 1 |
| **HYPOTHETICAL EXPERIMENT (B) / TRIMMING (b) (Restriction to girls between 10 and 18 years old and height between 60 and 69 inches and to boys between 9 and 18 years and height between 58 to 72 inches)** | | | | | | | |
| Age (years) | 361 | 9 | 10 | 12 | 11 | 13 | 18 |
| Height (inches) | 361 | 58 | 62 | 65 | 64 | 67 | 72 |
| Parental smoking (0: no, 1: yes) | 361 | 0 | 0 | 17% | 0 | 0 | 1 |
| Male children (0: no, 1: yes) | 361 | 0 | 0 | 59% | 1 | 1 | 1 |
| **HYPOTHETICAL EXPERIMENT (C) / STRATIFIED MATCHING (c) (*cem* R package)** | | | | | | | |
| Age (years) | 273 | 8 | 10 | 12 | 11 | 13 | 19 |
| Height (inches) | 273 | 57 | 62 | 65 | 65 | 67 | 74 |
| Parental smoking (0: no, 1: yes) | 273 | 0 | 0 | 21% | 0 | 0 | 1 |
| Male children (0: no, 1: yes) | 273 | 0 | 0 | 51% | 1 | 1 | 1 |
| **HYPOTHETICAL EXPERIMENTS (D.1 and D.2) / PROPENSITY SCORE MATCHING (d) (caliper=1 standard deviation of the propensity score, *Matching* R package)** | | | | | | | |
| Age (years) | 126 | 9 | 12 | 13 | 13 | 15 | 19 |
| Height (inches) | 126 | 58 | 64 | 67 | 66 | 69 | 74 |
| Parental smoking (0: no, 1: yes) | 126 | 0 | 0 | 50% | 0 | 1 | 1 |
| Male children (0: no, 1: yes) | 126 | 0 | 0 | 45% | 0 | 1 | 1 |
| **HYPOTHETICAL EXPERIMENT (E) / OPTIMAL PAIR MATCHING (e) (Minimum squared Mahalanobis distance, *optmatch* R package)** | | | | | | | |
| Age (years) | 126 | 9 | 12 | 13 | 13 | 15 | 18 |
| Height (inches) | 126 | 58 | 64 | 66 | 66 | 68 | 72 |
| Parental smoking (0: no, 1: yes) | 126 | 0 | 0 | 50% | 0 | 1 | 1 |

| Male children (0: no, 1: yes) | 126 | 0 | 0 | 41% | 0 | 1 | 1 |

Table 2: Assessing balance across design stage methods: mean (standard deviation) of the background covariates among children with smoking parents vs. children with non-smoking parents

| Hypothetical experiment / Design stage methods | Number of children | Average age | | Average height | | Male children proportion | |
|---|---|---|---|---|---|---|---|
| | | Children with smoking parents | Children with non-smoking parents | Children with smoking parents | Children with non-smoking parents | Children with smoking parents | Children with non-smoking parents |
| **HYPOTHETICAL EXPERIMENT (A) / NO DESIGN (a)** | 654 | 13.5 (2.34) | 9.5 (2.74) | 66.0 (3.19) | 60.6 (5.67) | 40% | 53% |
| **HYPOTHETICAL EXPERIMENT (B) / TRIMMING (b)** (Restriction to girls between 10 and 18 years old and height between 60 and 69 inches and to boys between 9 and 18 years and height between 58 to 72 inches) | 361 | 13.4 (2.17) | 11.4 (1.94) | 65.9 (3.24) | 64.3 (3.36) | 42% | 63% |
| **HYPOTHETICAL EXPERIMENT (C) / STRATIFIED MATCHING (c)** (*cem* R package) | 273 | 13.3 (2.32) | 11.6 (2.13) | 66.0 (3.09) | 64.6 (3.99) | 43% | 53% |
| **HYPOTHETICAL EXPERIMENTS (D.1 and D.2) / PROPENSITY SCORE MATCHING (d)** (caliper=1 standard deviation of the propensity score, *Matching* R package) | 126 | 13.5 (2.34) | 13.4 (2.31) | 66.0 (3.19) | 67.1 (3.89) | 40% | 49% |

| **HYPOTHETICAL EXPERIMENT (E) / OPTIMAL PAIR MATCHING (e) (Minimum squared Mahalanobis distance, *optmatch* R package)** | 126 | 13.3 (2.27) | 13.3 (2.16) | 66.0 (3.20) | 66.0 (3.24) | 41% | 41% |

Table 3: Analysis stage: comparison of the average causal effect (ACE) estimates and intervals across methods

| Hypothetical experiment / Design stage methods | Analysis method | Number of units | Estimate of the average causal effect (ACE) | 95% confidence interval |
|---|---|---|---|---|
| **HYPOTHETICAL EXPERIMENT (A) / NO DESIGN (a)** | Crude comparison | 654 | 0.71 | [0.50; 0.93] |
| | Standard linear regression with no interactions | 654 | -0.09 | [-0.20; 0.03] |
| **HYPOTHETICAL EXPERIMENT (B) / TRIMMING (b) (Restriction to girls between 10 and 18 years old and height between 60 and 69 inches and to boys between 9 and 18 years and height between 58 to 72 inches)** | Crude comparison | 361 | 0.18 | [-0.03; 0.39] |
| | Standard linear regression with no interactions | 361 | -0.16 | [-0.30; -0.03] |
| **HYPOTHETICAL EXPERIMENT (C) / STRATIFIED MATCHING (c) (*cem* R package)** | Crude comparison | 273 | -0.16 | [-0.37; 0.05] |
| | Standard linear regression with no interactions | 273 | -0.16 | [-0.30; -0.03] |
| **HYPOTHETICAL EXPERIMENTS (D.1 and D.2) / PROPENSITY SCORE MATCHING (d) (caliper=1 standard deviation of the propensity score, *Matching* R package)** | Crude comparison | 126 | -0.20 | [-0.43; 0.03] |
| | Standard linear regression with no interactions | 126 | -0.23 | [-0.46; -0.00] |
| **HYPOTHETICAL EXPERIMENT (E) / OPTIMAL PAIR MATCHING (e) (Mimimum squared Mahalanobis distance, *optmatch* R package)** | Crude comparison | 126 | -0.19 | [-0.46; 0.08] |
| | Standard linear regression with no interactions | 126 | -0.18 | [-0.35; -0.01] |

1 **FIGURES**

2 **Figure 1**

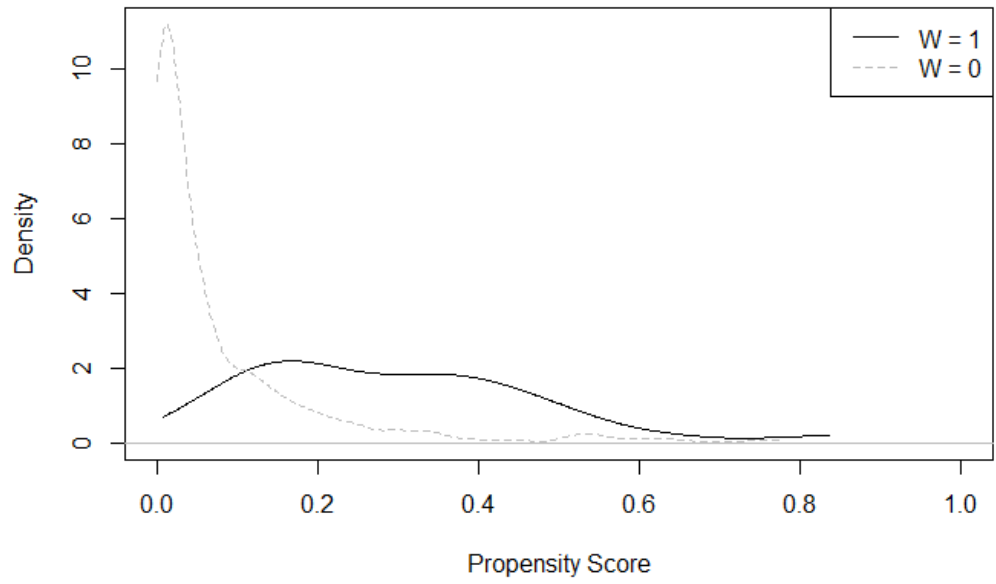

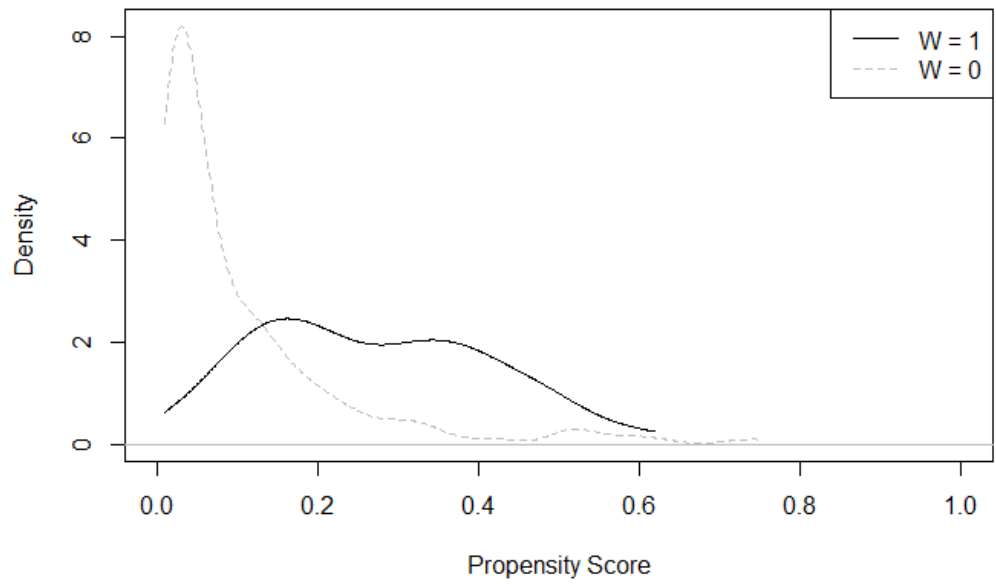

3
4

1 **Figure 2**

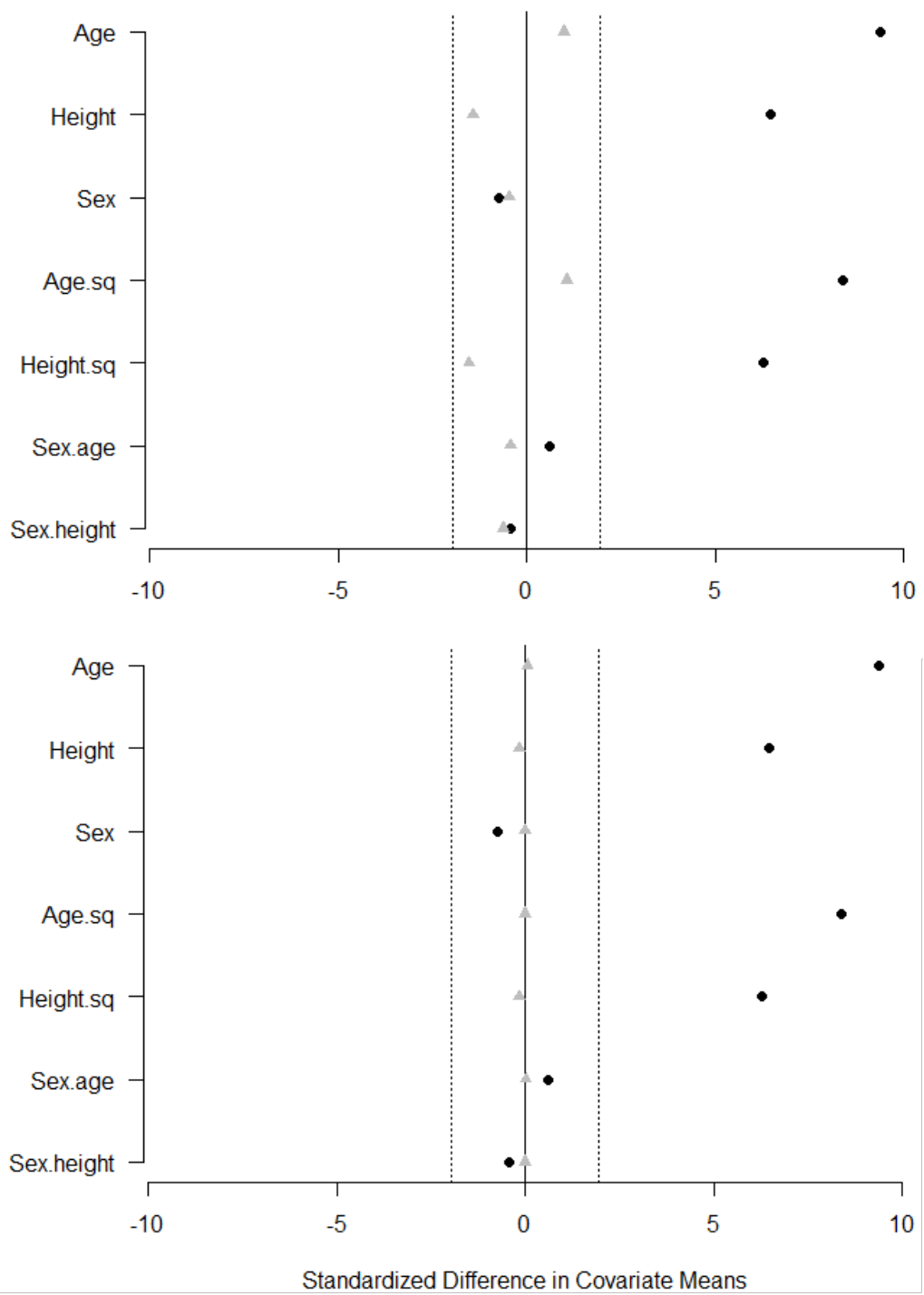

1 **Figure 3**

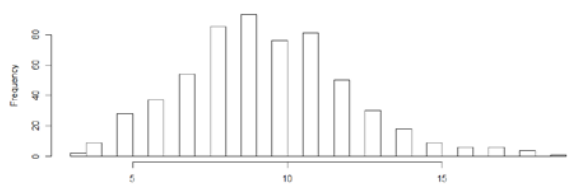 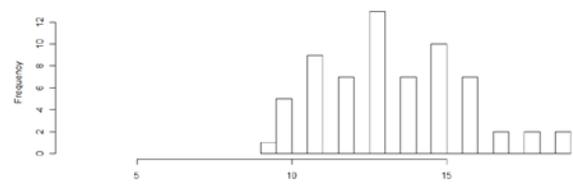

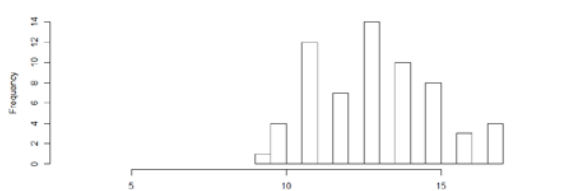 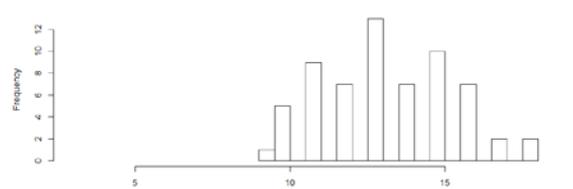

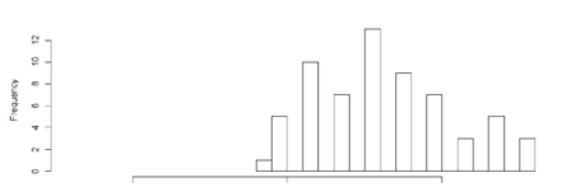 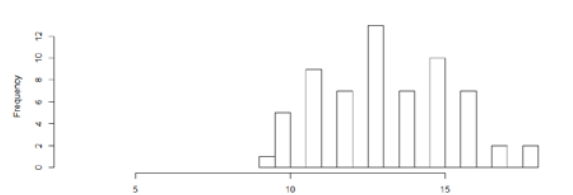

2
3

1 **Figure 4**

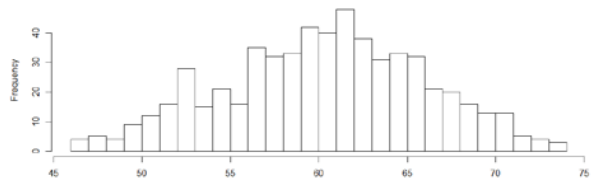
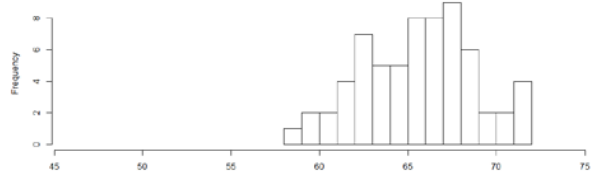
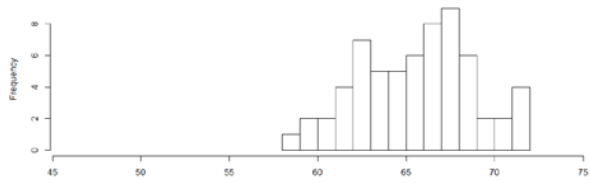
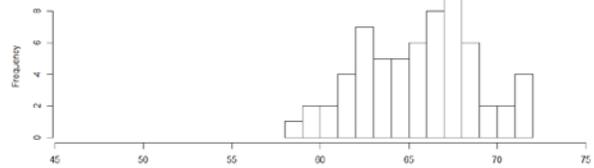
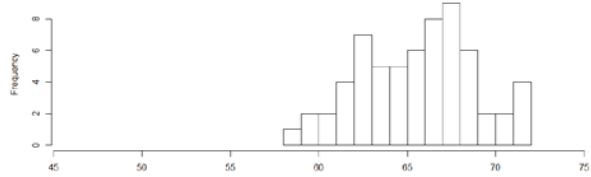
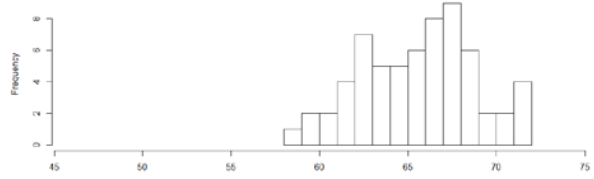

1  **Figure 5**

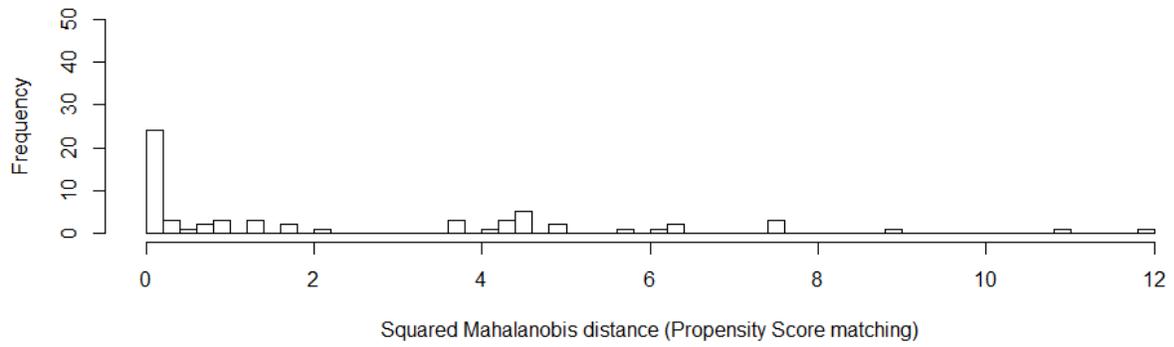

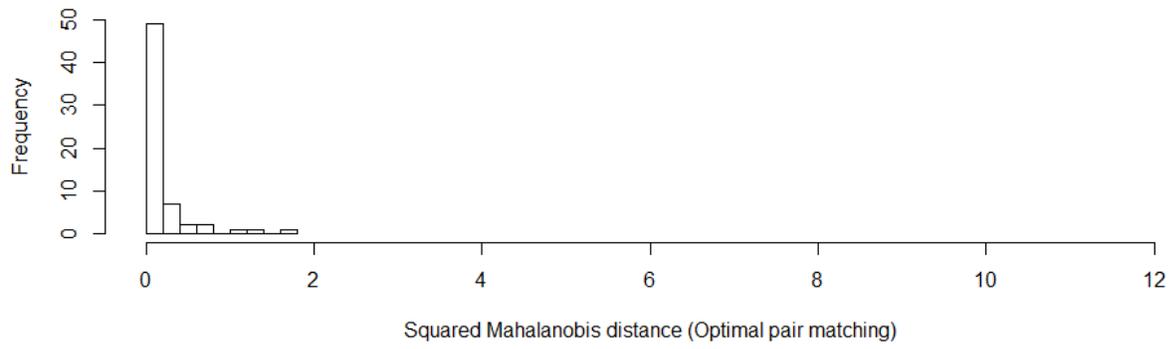

2
3

1 **Figure 6**

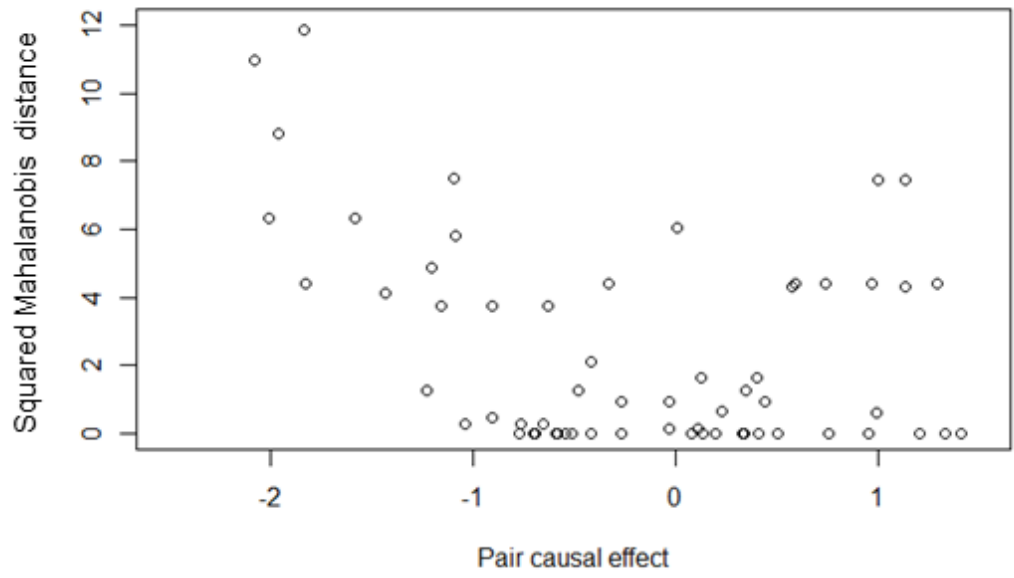

2
3

1    **Figure 7**

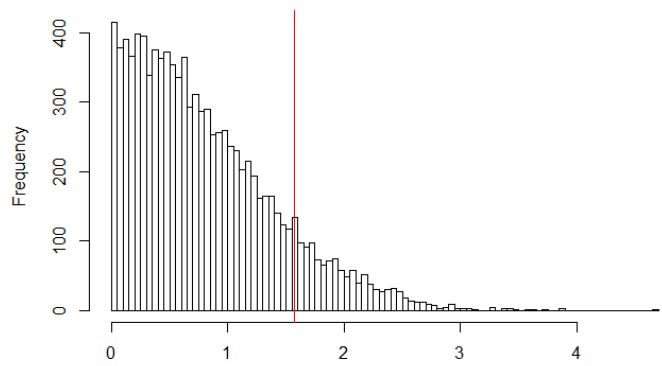

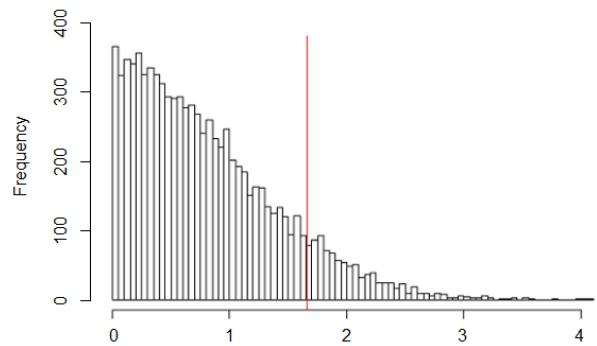

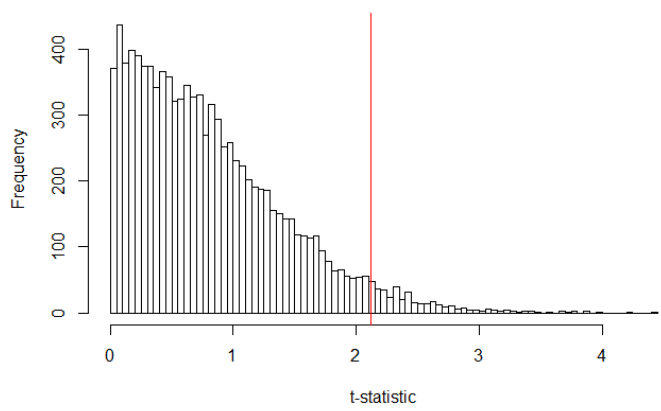

2
3

1   **Figure 8**

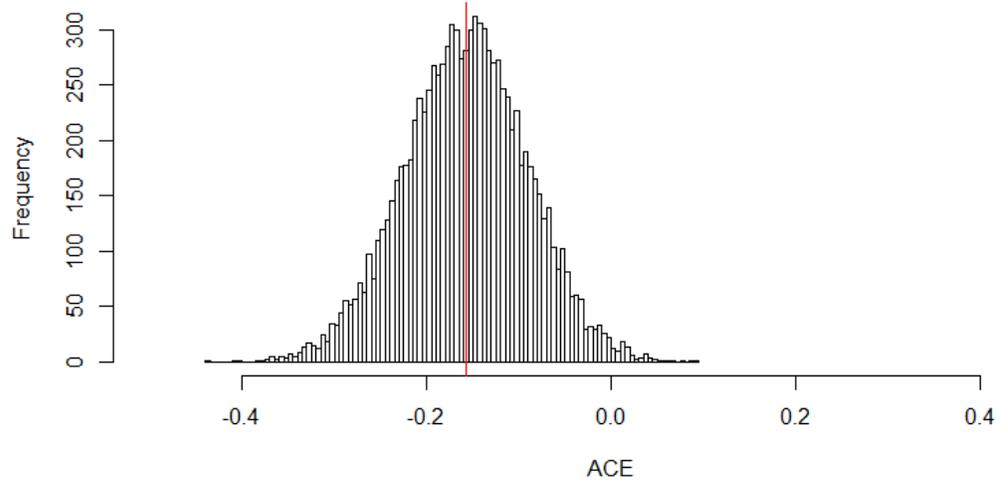

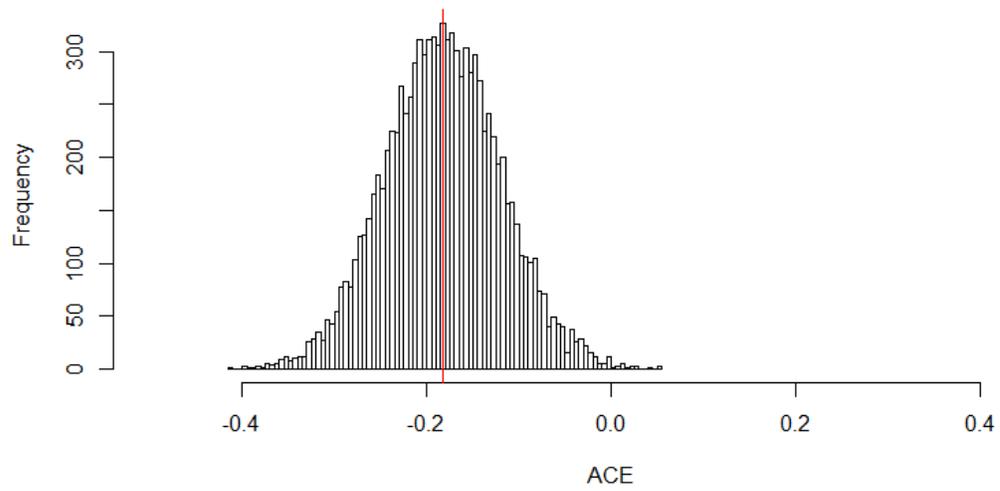

2
3
44

Figure 9

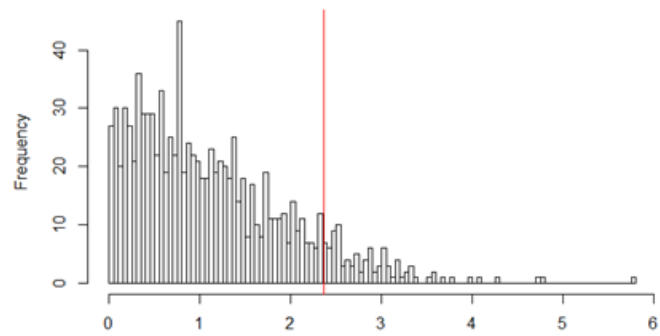

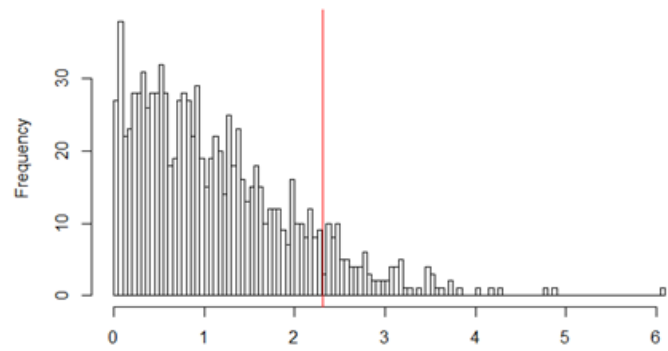

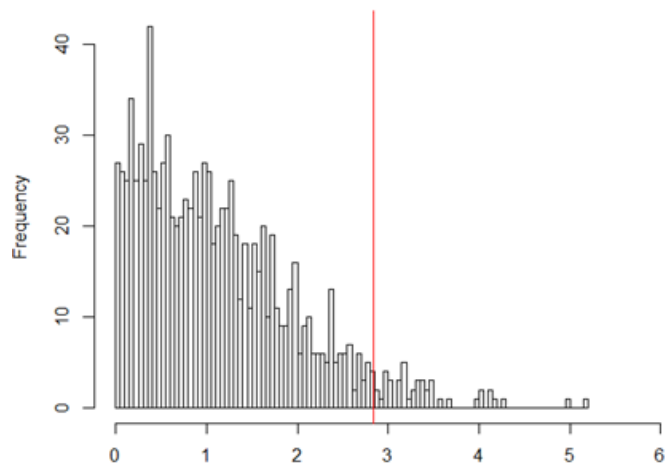

1  **SUPPLEMENTAL FIGURES**

2  **Supplemental Figure 1: Boxplots of FEV-1 distributions among girls (left boxplots) and**
3  **boys (right boxplots) for children with non-smoking parents (top boxplots) and children**
4  **with smoking parents (bottom boxplots)**

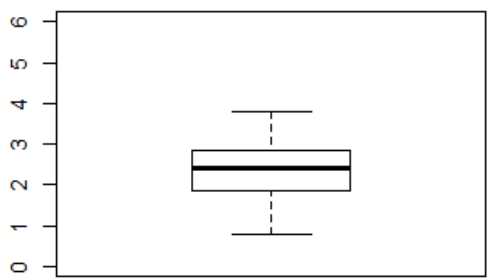
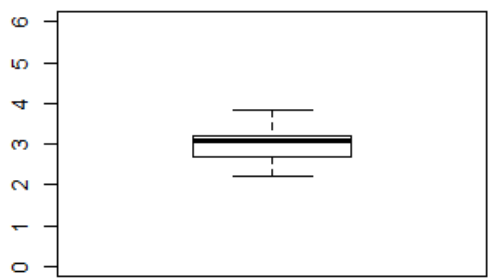
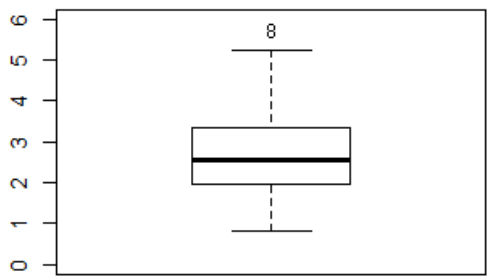
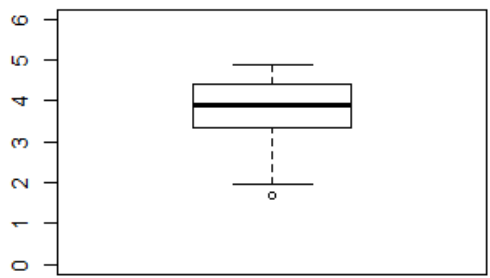

5
6

1   **Supplemental Figure 2: Estimated cubic splines for the FEV-1 vs. age and FEV-1 vs. height**

2   **relationships**

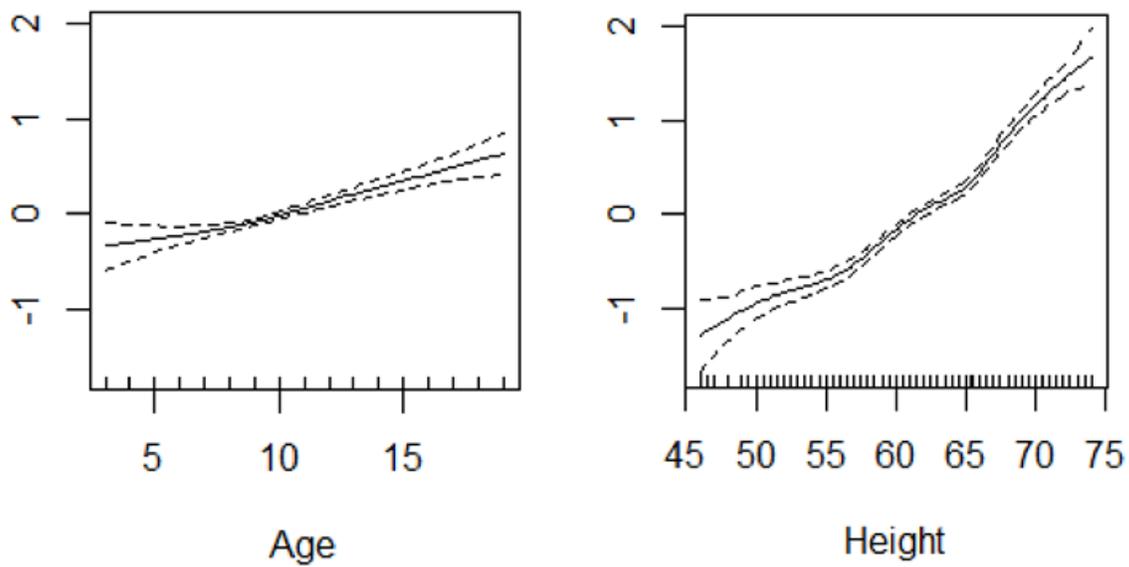

3
4
5